\documentclass[conference]{IEEEtran}

\usepackage{amsmath}
\usepackage{amssymb}
\usepackage{amsfonts}
\usepackage{graphicx}
\usepackage{epsfig}
\usepackage{subfigure}
\usepackage{psfrag}
\usepackage{cite}
\usepackage{latexsym}
\usepackage{url}
\usepackage{color}
\usepackage{multirow}

\PassOptionsToPackage{bookmarks={false}}{hyperref}

\begin{document}
\title{UAV-Enabled Wireless Power Transfer: Trajectory Design and Energy Region Characterization
}
\author{Jie Xu$^1$, Yong Zeng$^2$, and Rui Zhang$^2$\\
$^1$School of Information Engineering, Guangdong University of Technology\\
$^2$Department of Electrical and Computer Engineering, National University of Singapore\\
E-mail:~jiexu@gdut.edu.cn,~\{elezeng,~elezhang\}@nus.edu.sg

}

\maketitle

\begin{abstract}
This paper studies a new unmanned aerial vehicle (UAV)-enabled wireless power transfer (WPT) system, where a UAV-mounted energy transmitter (ET) broadcasts wireless energy to charge distributed energy receivers (ERs) on the ground. In particular, we consider a basic two-user scenario, and investigate how the UAV can optimally exploit its mobility to maximize the amount of energy transferred to the two ERs during a given charging period. We characterize the achievable energy region of the two ERs, by optimizing the UAV's trajectory subject to a maximum speed constraint. We show that when the distance between the two ERs is smaller than a certain threshold, the boundary of the energy region is achieved when the UAV hovers above a fixed location between them for all time; while when their distance is larger than the threshold, to achieve the boundary of the energy region, the UAV in general needs to hover and fly between two different locations above the line connecting them. Numerical results show that the optimized UAV trajectory can significantly improve the WPT efficiency and fairness of the two ERs, especially when the UAV's maximum speed is large and/or the charging duration is long.
\end{abstract}

\newtheorem{definition}{\underline{Definition}}[section]
\newtheorem{fact}{Fact}
\newtheorem{assumption}{Assumption}
\newtheorem{theorem}{\underline{Theorem}}[section]
\newtheorem{lemma}{\underline{Lemma}}[section]
\newtheorem{corollary}{\underline{Corollary}}[section]
\newtheorem{proposition}{\underline{Proposition}}[section]
\newtheorem{example}{\underline{Example}}[section]
\newtheorem{remark}{\underline{Remark}}[section]
\newtheorem{algorithm}{\underline{Algorithm}}[section]
\newcommand{\mv}[1]{\mbox{\boldmath{$ #1 $}}}
\setlength\abovedisplayskip{4pt}
\setlength\belowdisplayskip{4pt}

\section{Introduction}

Radio frequency (RF) transmission enabled wireless power transfer (WPT) has been regarded as a promising technique to provide perpetual and cost-effective energy supplies to low-power wireless networks (see, e.g., \cite{BiHoZhang2015,ZengZhang2017} and the references therein). In conventional WPT systems, dedicated energy transmitters (ETs) are usually deployed at fixed locations to charge distributed energy receivers (ERs) such as low-power sensors and Internet-of-things (IoT) devices, etc.

However, due to the severe propagation loss of RF signals over distance, the performance of practical WPT systems is constrained by the low end-to-end WPT efficiency and the short power coverage range. As a result, in order to provide ubiquitous wireless energy supply to massive low-power ERs, such fixed ETs need to be deployed in an ultra-dense manner. This, however, would tremendously increase the cost, and hinder the large-scale implementation of WPT systems. In the literature, different approaches have been proposed aiming to resolve this issue by enhancing the WPT efficiency at the {\it link level}, including e.g. multi-antenna beamforming \cite{XuLiuZhang2014,XuZhang2014,ZengZhang2015} and waveform optimization \cite{ClerckxBayguzina2016}. Different from the prior studies, in this paper we tackle this problem more cost-effectively from a new design perspective at the {\it system level}, and propose a radically novel architecture for WPT systems by utilizing unmanned aerial vehicles (UAVs) as mobile ETs.

\begin{figure}
\centering
 \epsfxsize=1\linewidth
    \includegraphics[width=9cm]{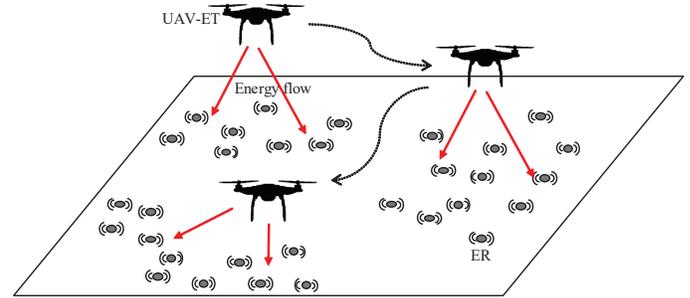}
\caption{Illustration of the UAV-enabled WPT system.} \label{fig:UAV:WPT}\vspace{-1.5em}
\end{figure}

UAV has recently drawn significant interests in many applications, such as weather monitoring, forest fire detection, communication relaying, etc. Particularly, by mounting communication transceivers to low-altitude UAVs, they can be used as aerial mobile base stations or relays to help enhance the performance of terrestrial wireless communication systems (see, e.g., \cite{ZengZhangLim2016} and the references therein). By optimally designing the UAV's trajectory and employing distance-based user scheduling, the communication link distance between the UAV and each of its served ground users can be effectively shortened, thus significantly improving the system throughput \cite{ZengZhangLim2016b}.

Motivated by UAV-assisted wireless communications, in this paper we study a new UAV-enabled WPT system as illustrated in Fig.~\ref{fig:UAV:WPT}. Specifically, with the proposed architecture, a group of UAVs are deployed as ETs that fly above a large area to cooperatively charge distributed ERs on the ground. By exploiting the fully controllable mobility introduced by UAVs via proper trajectory design, the new system is expected to significantly improve the WPT efficiency while reducing the number of required ETs as compared to the conventional WPT system with ETs deployed at fixed locations on the ground. Notice that there have been some prior works (e.g., \cite{Xie2012,Shu2016}) that proposed to use moving ground vehicles as mobile charging stations to wirelessly charge sensor nodes. Different from ground vehicles that can only move along a set of fixed trajectories in a two-dimensional (2D) area, UAVs can be more flexibly deployed and moved in the three-dimensional (3D) free space. Furthermore, compared to terrestrial wireless channels that typically suffer from various impairments such as shadowing and fading in addition to path loss, UAVs usually possess better channels to ground ERs due to the higher chance of having line-of-sight (LOS) links with them.

A fundamental question to be addressed in UAV-enabled WPT systems is as follows: how to jointly design the optimal trajectories of multiple UAVs to maximize the energy transferred to all ERs in a fair manner? This question, however, has not yet been studied in the literature to our best knowledge, and it is also non-trivial even for the simplest case with one UAV and two ERs. Notice that at each time in this case, the transferred powers from the UAV to the two ERs critically depend on the UAV's location, and when the UAV moves from one ER to the other, the received power will decrease/increase at the two ERs, respectively, thus resulting in a power trade-off between them.


To gain the most essential insights on the optimal UAV trajectory design, in this paper we focus on the basic UAV-enabled WPT system with two users and characterize their achievable energy region over a given charging period, which constitutes all the achievable energy pairs of the two ERs over all possible UAV trajectories subject to maximum speed constraints. In order to reveal the optimal performance trade-off between the two ERs, we aim to characterize the Pareto boundary of the energy region and their corresponding optimal UAV trajectory design. To this end, we adopt an {\it energy-profile} technique as in \cite{XuBiZhang2015}, and maximize the total energy transferred to the two ERs, subject to different fairness constraints between them.

First, we consider the ideal case by ignoring the UAV's maximum speed constraint, under which we find the optimal UAV trajectory to solve the energy-profile constrained total energy maximization problem. Based on this solution, we then propose an efficient solution for this problem in the general case with the UAV's speed limit considered. It is shown that when the distance between the two ERs is smaller than a certain threshold, the boundary of the energy region is achieved when the UAV hovers above a fixed location between them during the whole charging period; while when their distance is larger than the threshold, to achieve the boundary of the energy region, the UAV in general needs to hover and fly between two different locations above the line connecting the two ERs. Numerical results show that the optimized UAV trajectory can significantly improve the WPT efficiency as well as the performance fairness of the two ERs, especially with large values of  the UAV's maximum speed and/or the charging duration.

\section{System Model}

We consider a two-user UAV-enabled WPT system, where a UAV broadcasts wireless energy to charge two separated ERs on the ground. We consider a given charging period with duration $T$, denoted by $\mathcal{T} \triangleq [0,T]$. Each ER $k \in\{1,2\}$ has a fixed location on the ground, denoted by $(x_k,0)$ in 2D. Let $D$ denote the distance between the two ERs, then without loss of generality we assume $x_1 = -D/2$ and $x_2 = D/2$. The UAV is assumed to fly at a fixed altitude $H > 0$, whose time-varying location is denoted as $(x(t),H)$, $t \in \mathcal T$. Denote by $V$ in meter/second (m/s) the maximum possible speed of the UAV. We then have the constraint $|\dot{x}(t)| \le V, \forall t\in\mathcal T$, with $\dot{x}(t)$ denoting the time-derivative of $x(t)$.

As the wireless channel between the UAV and each ER is normally LOS-dominated, we adopt the free-space path loss model similarly as in  \cite{ZengZhangLim2016,ZengZhangLim2016b}. At time $t \in \mathcal T$, the channel power gain from the UAV to ER $k \in \{1,2\}$ is modeled as $h_k(t) = \beta_0 d_k^{-2}(t)$, where $d_k(t) = \sqrt{(x(t) - x_k)^2 + H^2}$ is their distance and $\beta_0$ denotes the channel power gain at a reference distance of $d_0 = 1$ m. Assuming that the UAV has a constant transmit power $P$, the harvested power by ER $k$ at time $t$ is thus given by
\begin{align}\label{eqn:harvested:power}
Q_k(x(t)) = \eta h_k(t)P = \frac{\eta\beta_0P}{(x(t) - x_k)^2 + H^2},
\end{align}
where $0 < \eta <1$ denotes the energy conversion efficiency of the rectifier at each ER. Thus, the total energy harvested by each ER $k \in \{1,2\}$ over the total duration is a function of the UAV trajectory $\{x(t)\}$, which can be written as
\begin{align}\label{eqn:harvested:energy}
E_k(\{x(t)\}) = \int_{0}^T Q_k(x(t)) \text{d}t.
\end{align}

Next, we define the achievable energy region for the two-user UAV-enabled WPT system, which constitutes all the achievable energy pairs of the two ERs under all feasible UAV trajectories $\{x(t)\}$ satisfying the maximum speed constraint, i.e.,
\begin{align}
&\mathcal{E} = \nonumber\\
&\bigcup_{|\dot{x}(t)| \le V, \forall t\in\mathcal T}\{(e_1,e_2): 0\le e_k \le E_k(\{x(t)\}), \forall k\in\{1,2\}\}.
\end{align}
Note that this region is generally not a convex set, due to the non-concavity of $E_k(\{x(t)\})$ with respect to $\{x(t)\}$. Our objective is thus to characterize the {\it Pareto boundary} of the energy region, which consists of all energy pairs at each of which it is impossible to improve one ER's harvested energy without simultaneously decreasing the other's. Towards this end, we maximize the total energy transferred to both ERs, subject to different energy fairness constraints based on the technique of {\it energy-profile} \cite{XuBiZhang2015}.\footnote{Another commonly adopted approach to characterize the Pareto boundary of the energy region is via maximizing the weighted sum of the harvested energy of the ERs with different user weights. However, as the energy region $\mathcal{E}$ is generally a non-convex set, this approach may fail to characterize all the Pareto boundary points (see, e.g.,  \cite{ZhangCui2010} and \cite[Chapter 4.7.3]{Boyd:Book}).} Mathematically, we formulate the following optimization problem with a particular energy-profile vector $\mv{\alpha} = [\alpha_1,\alpha_2]$:
\begin{align}
\text{(P1)}:\max_{\{x(t)\},E} ~& E \nonumber\\
\mathrm{s.t.}~&
\int_{0}^T {Q}_k(x(t)) \mathrm{d} t \ge \alpha_k E, \forall k\in\{1,2\}\label{eqn:con:1} \\
~
& |{\dot{x}}(t)| \le {V}, \forall t \in \mathcal T\label{eqn:con:2},
\end{align}
where the variable $E$ denotes the total energy harvested by the two ERs, and the constraints in (\ref{eqn:con:1}) specify the energy fairness between the two ERs. Here, $\alpha_k\ge 0, k\in\{1,2\}$, denotes the target ratio of ER $k$'s harvested energy over $E$ with $\alpha_1 + \alpha_2 = 1$.

Note that problem (P1) is difficult to solve optimally, since it involves an infinite number of variables $\{x(t)\}$ and the constraints in (\ref{eqn:con:1}) are non-convex due to the non-concavity of ${Q}_k(x(t))$. To tackle this difficulty, in Section \ref{sec:P2} we first consider the ideal case without the speed constraints in (\ref{eqn:con:2}) and solve this problem optimally. Note that the speed constraints in (\ref{eqn:con:2}) can be approximately ignored in practice if the product of the charging duration $T$ and the maximum UAV speed $V$ is much larger than the distance $D$ between the two ERs. For ease of presentation, we rewrite problem (P1) without (\ref{eqn:con:2}) in the following problem denoted by (P2).
\begin{align}
\text{(P2)}:\max_{\{x(t)\},E} ~& E \nonumber\\
\mathrm{s.t.}~&\eqref{eqn:con:1}\nonumber
\end{align}
In Section \ref{sec:P1}, we proceed to consider the general case of (P1) with the UAV speed constraints in (\ref{eqn:con:2}), and propose an efficient solution to (P1) based on the optimal solution obtained for (P2).

%
%

\section{Optimal Solution to Problem (P2)}\label{sec:P2}

Though problem (P2) is still non-convex, it can be easily shown that it satisfies the so-called time-sharing condition \cite{YuLui2006}. Therefore, the strong duality holds between (P2) and its dual problem. As a result, we can solve (P2) by using the Lagrange dual method \cite{Boyd:Book}.

Let $\lambda_k \ge 0$, $k\in\{1,2\}$, denote the dual variable associated with the $k$th constraint in \eqref{eqn:con:1}. The Lagrangian of (P2) is thus
\begin{align}
\mathcal{L}(\{x(t)\},E,\lambda_1,\lambda_2) = E + \sum_{k=1}^2 \lambda_k \left(\int_{0}^T Q_k(x(t)) \mathrm{d} t - \alpha_k E\right).
\end{align}
Accordingly, the dual function of (P2) is
\begin{align}\label{eqn:dual}
f(\lambda_1,\lambda_2) = \max_{\{x(t)\},E}\mathcal{L}(\{x(t)\},E,\lambda_1,\lambda_2),
\end{align}
for which the following lemma holds.
\begin{lemma}\label{lemma:1}
In order for $f(\lambda_1,\lambda_2)$ to be bounded from above (i.e., $f(\lambda_1,\lambda_2) < \infty$), it must hold that $\sum_{k=1}^2 \alpha_k\lambda_k = 1$.
\end{lemma}
\begin{IEEEproof}
Suppose that $\sum_{k=1}^2 \alpha_k\lambda_k> 1$ (or $\sum_{k=1}^2 \lambda_k\alpha_k <1$). Then by setting $E \to -\infty$ (or $E \to \infty$), we have $f(\lambda_1,\lambda_2)\to \infty$. Therefore, this lemma is proved.
\end{IEEEproof}
Based on Lemma \ref{lemma:1}, the dual problem of (P2) is given by
\begin{align}
\text{(D2)}:\min_{\lambda_1\ge 0,\lambda_2\ge 0}&f(\lambda_1,\lambda_2) \nonumber\\
{\text{s.t.}}~&\sum_{k=1}^2 \alpha_k\lambda_k = 1.\label{eqn:D2}
\end{align}
Then, we can solve problem (P2) by equivalently solving its dual problem (D2). Let the feasible set of $\lambda_1$ and $\lambda_2$ characterized by $\lambda_1\ge 0$, $\lambda_2\ge 0$, and $\sum_{k=1}^2 \alpha_k\lambda_k = 1$ as $\mathcal X$. In the following, we first solve problem (\ref{eqn:dual}) to obtain $f(\lambda_1,\lambda_2)$ under any given $(\lambda_1 ,\lambda_2)\in \mathcal X$, and then find the optimal $\lambda_1$ and $\lambda_2$ to minimize $f(\lambda_1,\lambda_2)$.

\subsection{Obtaining $f(\lambda_1,\lambda_2)$ by Solving Problem (\ref{eqn:dual})}

For any given $(\lambda_1 ,\lambda_2)\in \mathcal X$, problem \eqref{eqn:dual} can be decomposed into different subproblems as follows.
\begin{align}
\max_{x(t)}~ & \psi_{\lambda_1,\lambda_2}(x(t)) \triangleq \sum_{k=1}^2 \lambda_k Q_k(x(t)),~\forall t\in\mathcal T \label{eqn:sub:1}\\
\max_{E} ~ & \bigg(1 - \sum_{k=1}^2 \alpha_k\lambda_k\bigg)E \label{eqn:sub:2}
\end{align}
Here, (\ref{eqn:sub:1}) consists of an infinite number of subproblems, each corresponding to a time instant $t$. Let the optimal solutions to \eqref{eqn:sub:1} and \eqref{eqn:sub:2} be denoted by $x_{\lambda_1,\lambda_2}^*(t)$'s, $\forall t\in\mathcal T$, and $E_{\lambda_1,\lambda_2}^*$, respectively. Note that each subproblem in (\ref{eqn:sub:1}) is irrespective of the time index $t$; thus in this subsection we can denote the function $\psi_{\lambda_1,\lambda_2}(x(t))$ and the solution $x_{\lambda_1,\lambda_2}^*(t)$ as $\psi_{\lambda_1,\lambda_2}(x)$ and $x_{\lambda_1,\lambda_2}^*$, respectively, by dropping the index $t$.

\begin{figure*}
\begin{align}\label{eqn:first:order}
\psi_{\lambda_1,\lambda_2}'(x) 
&= - \eta\beta_0 P \frac{\lambda_1(2x(t)+D)(x^2+ D^2/4 + H^2 -Dx)^2 + \lambda_2(2x-D)(x^2 + D^2/4 + H^2 + Dx)^2 }{(x^2+ D^2/4 + H^2 -Dx)^2(x^2 + D^2/4 + H^2 + Dx)^2}.
\end{align}\vspace{-2em}
\end{figure*}
As for problem (\ref{eqn:sub:2}), since $1-\sum_{k=1}^2 \alpha_k\lambda_k =0$ holds for any given $(\lambda_1 ,\lambda_2)\in \mathcal X$, the objective value is always zero. In this case, we can choose any arbitrary real number as the optimal solution $E_{\lambda_1,\lambda_2}^*$ for the purpose of obtaining the dual function $f(\lambda_1,\lambda_2)$.

Next, we only need to consider problem (\ref{eqn:sub:1}) under $(\lambda_1 ,\lambda_2)\in \mathcal X$. However, the objective function $\psi_{\lambda_1,\lambda_2}(x)$ in (\ref{eqn:sub:1}) is non-convex, thus making this problem difficult to solve. Fortunately, there is only one single variable in (\ref{eqn:sub:1}). Therefore, it can be solved by first deriving the first-order derivative of $\psi_{\lambda_1,\lambda_2}(x)$, denoted by $\psi_{\lambda_1,\lambda_2}'(x)$ given in (\ref{eqn:first:order}) at the top of next page, and then comparing all the solutions to the equation $\psi_{\lambda_1,\lambda_2}'(x) = 0$. As the numerator in the right-hand-side (RHS) of (\ref{eqn:first:order}) is a polynomial function with the highest power being five, it is evident that there are at most five real solutions to $\psi_{\lambda_1,\lambda_2}'(x) = 0$. By comparing all these solutions, the optimal solution $x_{\lambda_1,\lambda_2}^*$ to problem (\ref{eqn:sub:1}) can be obtained as the one with the largest objective value. If there are more than one solution achieving the same largest objective value, then we can arbitrarily choose any one of them as $x_{\lambda_1,\lambda_2}^*$.\footnote{Note that the optimal solution $E_{\lambda_1,\lambda_2}^*$ and $x_{\lambda_1,\lambda_2}^*$ are not unique in general, and thus they may not be feasible for the primal problem (P2). As a result, an additional step is required to obtain the primal feasible and optimal solution of $E$ and $x(t)$'s to (P2), as will be shown in Section \ref{sec:construct} later.} By substituting $x_{\lambda_1,\lambda_2}^*$ to problem (\ref{eqn:sub:1}) and $E_{\lambda_1,\lambda_2}^*$ to problem (\ref{eqn:sub:2}), the function $f(\lambda_1,\lambda_2)$ is finally obtained.

In the following, we provide insights on the optimal solution $x_{\lambda_1,\lambda_2}^*$ to problem (\ref{eqn:sub:1}). When $\lambda_1 = \lambda_2$, we have the closed-form solution to (\ref{eqn:sub:1}) as follows.


\setlength\abovedisplayskip{1pt}
\setlength\belowdisplayskip{1pt}

\begin{lemma}\label{lemma:3.2}
In the case with $\lambda_1 = \lambda_2 = \lambda$, the optimal solution to problem (\ref{eqn:sub:1}) is given by
\begin{align}
&x^*_{\lambda,\lambda} \triangleq
\left\{\begin{array}{ll}
\pm \xi, & {\text{if}}~D > 2H/\sqrt{3},\\
0,& {\text{if}}~D \le 2H/\sqrt{3},
\end{array}\right.
\end{align}
where
\begin{align}\label{eqn:xi}
\xi \triangleq \sqrt{-(D^2/4+H^2) + \frac{\sqrt{D^4/4 + H^2D^2}}{2}} < D/2.
\end{align}
In other words, if $D > 2H/\sqrt{3}$, then problem (\ref{eqn:sub:1}) has two optimal solutions; while if $D \le 2H/\sqrt{3}$, then problem (\ref{eqn:sub:1}) has only one unique optimal solution, which is zero.
\end{lemma}
\begin{IEEEproof}
See Appendix \ref{appendix:A}.
\end{IEEEproof}

When $\lambda_1 \neq \lambda_2$, we have the following remark.
\begin{remark}\label{remark:1}
In the case with $\lambda_1 \neq \lambda_2$, the optimal solution $x^*_{\lambda_1,\lambda_2}$ to problem (\ref{eqn:sub:1}) is unique, which satisfies
\begin{align}
x^*_{\lambda_1,\lambda_2} \in
\left\{\begin{array}{ll}
\left[-D/2,-\xi\right], & {\text{if}}~D > 2H/\sqrt{3}~{\text{and}}~\lambda_1 > \lambda_2,\\
\left[-D/2,0\right],& {\text{if}}~D \le 2H/\sqrt{3}~{\text{and}}~\lambda_1 > \lambda_2,\\
\left[0,D/2\right],& {\text{if}}~D \le 2H/\sqrt{3}~{\text{and}}~\lambda_1 < \lambda_2,\\
\left[\xi,D/2\right],& {\text{if}}~D > 2H/\sqrt{3}~{\text{and}}~\lambda_1 < \lambda_2.\end{array}
\right.
\end{align}
\end{remark}

Note that Remark \ref{remark:1} is validated via extensive simulations with different values of $D$ and $H$, although it is difficult to be rigorously proved due to the complication of the first-order derivative $\psi'_{\lambda_1,\lambda_2}(x)$ in \eqref{eqn:first:order}. Lemma \ref{lemma:3.2} and Remark \ref{remark:1} are essential to draw insights on the optimal solution to the primal problem (P2) later.


%
%
%
%
%
%
%
%

\subsection{Finding Optimal Dual Solution to (D2)}\label{sec:dual}

With $f(\lambda_1,\lambda_2)$ obtained, we then solve the dual problem (D2) to find the optimal $\lambda_1$ and $\lambda_2$ to minimize $f(\lambda_1,\lambda_2)$. Note that the dual function $f(\lambda_1,\lambda_2)$ is always convex but generally non-differentiable \cite{Boyd:Book}. As a result, problem (D2) can be solved by subgradient based methods such as the ellipsoid method \cite{BoydII}. Note that the subgradient of the objective function $f(\lambda_1,\lambda_2)$ is \vspace{-1em}

\begin{small}
\begin{align*}
&\mv{s}_0(\lambda_1,\lambda_2) \nonumber\\
 = &\left[ \int_{0}^T Q_1(x^*_{\lambda_1,\lambda_2}) \mathrm{d} t - \alpha_1E_{\lambda_1,\lambda_2}^*, \int_{0}^T Q_2(x^*_{\lambda_1,\lambda_2}) \mathrm{d} t-\alpha_2E_{\lambda_1,\lambda_2}^*\right]^\ddagger \\
 =&\left[ T Q_1(x^*_{\lambda_1,\lambda_2}), T Q_2(x^*_{\lambda_1,\lambda_2})\right]^\ddagger,
\end{align*}
\end{small}where we choose $E_{\lambda_1,\lambda_2}^* = 0$ for simplicity, though we can also choose any other real numbers for $E_{\lambda_1,\lambda_2}^*$. Here, the superscript $\ddagger$ denotes the transpose. Furthermore, the equality constraint in (\ref{eqn:D2}) can be viewed as two inequality constraints $1 - \sum_{k=1}^2 \alpha_k\lambda_k \le 0$ and $-1 + \sum_{k=1}^2 \alpha_k\lambda_k \le 0$, whose subgradients are given by $\mv{s}_1(\lambda_1,\lambda_2) = [-\alpha_1,-\alpha_2]^\ddagger$ and $\mv{s}_2(\lambda_1,\lambda_2) = [\alpha_1,\alpha_2]^\ddagger$, respectively. We denote the obtained dual solution to (D2) as $\lambda_1^\star$ and $\lambda_2^\star$.

\subsection{Constructing Optimal Solution to (P2)}\label{sec:construct}

Based on the dual optimal solution $\lambda_1^\star$ and $\lambda_2^\star$ to (D2), we need to obtain the primal optimal solution to (P2), denoted by $\{x^\star(t)\}$ and $E^\star$. It is worth noting that when using the Lagrange dual method to solve problem (P2) via the dual problem (D2), the optimal solution to problem (\ref{eqn:dual}) under $\lambda_1^\star,\lambda_2^\star$ (i.e., $x^*_{\lambda_1^\star,\lambda_2^\star}(t)$ and $E_{\lambda_1^\star,\lambda_2^\star}^*$) is the primal optimal solution to (P2), if and only if such a solution is {\it unique and primal feasible} \cite{Boyd:Book}. On the other hand, when $x^*_{\lambda_1^\star,\lambda_2^\star}(t)$ and $E_{\lambda_1^\star,\lambda_2^\star}^*$ for problem \eqref{eqn:dual} are non-unique and infeasible to (P2), they are not the optimal solution to (P2) in general. In this case, we need some additional steps to construct the primal optimal solution $\{x^\star(t)\}$ and $E^\star$ to (P2).

We have the following two propositions.

\begin{proposition}\label{proposition:1}
If $\lambda_1^\star = \lambda_2^\star = \lambda^\star$, the optimal solution of $\{x^\star(t)\}$ and $E^\star$ to (P2) is given as follows:
\begin{itemize}
  \item In the case with $D \le 2H/\sqrt{3}$, we have $x^\star(t) = 0, \forall t \in \mathcal{T}$. Accordingly, the energies transferred to the two ERs are $E_1^\star = E_2^\star = TQ_1(0)= TQ_2(0)$, and we have $E^\star = 2TQ_1(0)$. This case only occurs when $\alpha_1 = \alpha_2 = 1/2$.
  \item In the case with $D > 2H/\sqrt{3}$, the charging period $\mathcal T$ is divided into two phases $\mathcal T_1 = [0,\tau]$ and $\mathcal T_2 = (\tau,T]$, and
  we have $x^\star(t) = -\xi, \forall t \in \mathcal T_1$, and $x^\star(t) = \xi, \forall t \in \mathcal T_2$, with $\xi$ given in (13). Accordingly, the energy transferred to each ER $k\in\{1,2\}$ is $E_k^\star = \tau Q_k(-\xi) + (T-\tau)Q_k(\xi)$. Here, $\tau$ is a constant that is chosen such that $\frac{E_1^\star}{E_2^\star} = \frac{\alpha_1}{\alpha_2}$. In this case, we have $E^\star = E_1^\star + E_2^\star.$
\end{itemize}
\end{proposition}
\begin{IEEEproof}
See Appendix \ref{appendix:B}.
\end{IEEEproof}

\begin{proposition}\label{proposition:2}
If $\lambda_1^\star \neq \lambda_2^\star$, the optimal solution to (P2) is given as $x^\star(t) = x^*_{\lambda_1^\star,\lambda_2^\star}, \forall t \in \mathcal{T}$. Accordingly, the energy transferred to each ER $k\in\{1,2\}$ is $E_k^\star = T Q_k(x^*_{\lambda_1^\star,\lambda_2^\star})$, and we have $E^\star = E_1^\star + E_2^\star.$
\end{proposition}
\begin{IEEEproof}
This proposition follows based on the fact that when $\lambda_1^\star \neq \lambda_2^\star$, the optimal solution $x^*_{\lambda_1^\star,\lambda_2^\star}$ to problem (\ref{eqn:sub:1}) is unique as shown in Remark \ref{remark:1}.
\end{IEEEproof}

Propositions \ref{proposition:1} and \ref{proposition:2} provide interesting structures for the optimal solution to problem (P2). If the distance between the two ERs is small (i.e., $D \le 2H/\sqrt{3}$), then the UAV only needs to hover above a fixed ground location to achieve the corresponding Pareto-optimal point, regardless of the values of $\lambda_1^\star$ and $\lambda_2^\star$. The ground location becomes the middle point between the two ERs when $\lambda_1^\star = \lambda_2^\star$. If the distance between the two ERs is large (i.e., $D > 2H/\sqrt{3}$), then in the case with $\lambda_1^\star = \lambda_2^\star$, the UAV needs to hover at two symmetric locations over the middle point of the two ERs successively, with one at $(-\xi,H)$ closer to ER 1 and the other at $(\xi,H)$ closer to ER 2, each for a certain portion of the total time; while in the case with $\lambda_1^\star \neq \lambda_2^\star$, the UAV only needs to hover at one fixed location to achieve the corresponding Pareto-optimal point.

%
%
%
%
%

%
%

%
%

\begin{figure}
\centering
 \epsfxsize=1\linewidth
    \includegraphics[width=7cm]{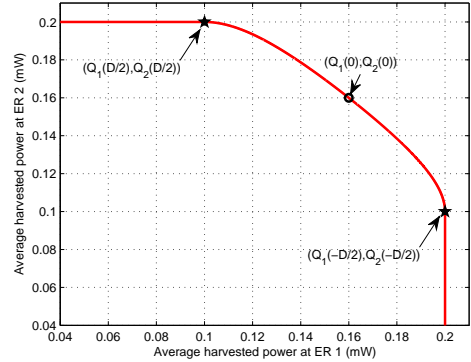}
\caption{Energy region of the two-user UAV-enabled WPT system, where the distance between the two ERs is $D = 5$ m.} \label{fig:3D:1}\vspace{-2em}
\end{figure}
\begin{figure}
\centering
 \epsfxsize=1\linewidth
    \includegraphics[width=7cm]{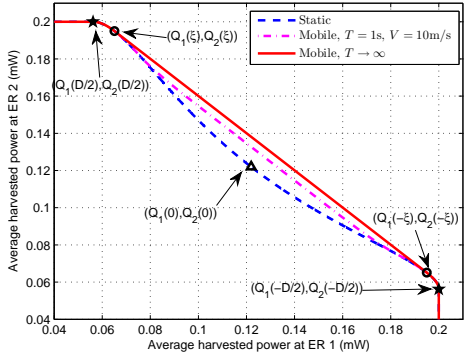}
\caption{Energy region of the two-user UAV-enabled WPT system, where the distance between the two ERs is $D = 8$ m.} \label{fig:3D:2}\vspace{-2em}
\end{figure}

\begin{example}\label{example:1}
For better illustration, Figs. \ref{fig:3D:1} and \ref{fig:3D:2} show the Pareto boundary of the energy region with $H=5$ m, $P = 40$ dBm, and $\eta = 50\%$. We consider two examples when $D = 5$ m $< 2H/\sqrt{3}$ and $D = 8$ m $> 2H/\sqrt{3}$, respectively. Here, the amount harvested energy at the two ERs is normalized by the charging duration $T$, and thus the average harvested power $\alpha_k E/T$ is considered for each ER $k\in\{1,2\}$. In both figures, the solid red curve corresponds to the Pareto boundary with the maximum speed constraints ignored (i.e., $V\to\infty$), which is obtained by optimally solving (P2) under different energy-profile values of $\mv{\alpha}$. We have the following observations. First, it is observed that for both cases with $D = 5$ m and $D = 8$ m, the energy regions of the two ERs are convex sets. This is due to the fact that problem (P2) satisfies the time-sharing condition and thus the strong duality holds between (P2) and its dual problem (D2). Next, it is also observed that when the UAV hovers at the location above ER 1 (with $x(t) = -D/2, \forall t \in \mathcal T$) or ER 2 (with $x(t) = D/2, \forall t \in \mathcal T$), the boundary point $(Q_1(-D/2),Q_2(-D/2))$ or $(Q_1(D/2),Q_2(D/2))$ is achieved, in which the energy transferred to ER 1 or ER 2 is maximized, respectively. Furthermore, when $D=5$ m, it is observed from Fig. \ref{fig:3D:1} that when the UAV hovers above the middle location between the two ERs (with $x(t) = 0, \forall t \in \mathcal T$), $(Q_1(0),Q_2(0))$ corresponds to one boundary point of the energy region. In this case, the boundary points between $(Q_1(0),Q_2(0))$ and $(Q_1(-D/2),Q_2(-D/2))$ (or $(Q_1(D/2),Q_2(D/2))$) are obtained via the UAV hovering at one fixed location between $(0,H)$ and $(-D/2,H)$ (or $(D/2,H)$). By contrast, when $D=8$ m, it is observed from Fig. \ref{fig:3D:2} that when the UAV hovers at the middle location between the two ERs, $(Q_1(0),Q_2(0))$ is not on the Pareto boundary of the energy region; while the boundary points between $(Q_1(-\xi),Q_2(-\xi))$ and $(Q_1(\xi),Q_2(\xi))$ are obtained via the UAV time-sharing between the two different hovering locations $(-\xi,H)$ and $(\xi,H)$. These observations are consistent with Propositions \ref{proposition:1} and \ref{proposition:2}.
\end{example}

%
%

\section{Proposed Solution to Problem (P1)}\label{sec:P1}

In this section, we consider problem (P1) in the general case when the UAV's maximum speed constraints in (\ref{eqn:con:2}) are taken into account. However, due to such constraints, it is difficult to use the Lagrange dual method to find the optimal solution to problem (P1). Thus, we solve (P1) based on the solution to (P2) obtained in the previous section. Let the obtained solution to (P1) be denoted by $\{x^{\star\star}(t)\}$ and $E^{\star\star}$. We first present the following proposition.
\begin{proposition}\label{proposition:3}
When $D \le 2H/\sqrt{3}$ or $\lambda_1^\star \neq \lambda_2^\star$ for (D2), the optimal solution to (P1) is identical to that to (P2), i.e., $x^{\star\star}(t) = x^\star(t), \forall t\in\mathcal T$, and $E^{\star\star}= E^\star$.
\end{proposition}
\begin{IEEEproof}
In the case with $D \le 2H/\sqrt{3}$ or $\lambda_1^\star \neq \lambda_2^\star$, it is evident that the optimal solution to problem (P2) is also feasible to (P1), as the UAV only needs to hover at one single location without violating the speed constraints. As the optimal value achieved by (P2) serves as an upper bound on that by (P1), the optimal solution to (P2) is also that to (P1).
\end{IEEEproof}

Next, it only remains to consider the case when $D > 2H/\sqrt{3}$ and $\lambda_1^\star = \lambda_2^\star$ for (D2), in which case the UAV needs to time-share between two hovering locations at the optimality of problem (P2) that ignores the UAV speed constraints, and thus the optimal solution of $\{x^\star(t)\}$ to (P2) is generally not feasible to (P1) with the UAV speed constraints considered. As a result, optimally solving (P1) becomes quite difficult in this case. In the following, we first solve (P1) when $\alpha_1 = \alpha_2 = 1/2$ to obtain the boundary point when the two ERs need to harvest the same amount of energy,{\footnote{Note that if $\alpha_1 = \alpha_2 = 1/2$, then we have $\lambda_1^\star = \lambda_2^\star$ for (D2). But the reverse is not true in general.}} and then propose an efficient method to obtain a suboptimal solution to (P1) when $\alpha_1 \neq \alpha_2$.

\subsection{Optimal Solution to (P1) when $D > 2H/\sqrt{3}$ and $\alpha_1 = \alpha_2 = 1/2$}

With $\alpha_1 = \alpha_2 = 1/2$, problem (P1) can be re-expressed as
\begin{align}
\text{(P3)}:& \max_{\{x(t)\},E} E \nonumber\\
\text{s.t.}~& \int_{0}^T Q_k(x(t)) \mathrm{d} t \ge E/2, \forall k\in\{1,2\} \nonumber\\
&|{\dot{x}}(t)| \le {V}, \forall t \in \mathcal T.
\end{align}

\begin{proposition}\label{proposition:4}
The optimal solution to (P3) has the following {\it hover-fly-hover} structure based on two symmetric locations $(-\hat{x},H)$ and $(\hat{x},H)$ (with $0 < \hat{x} \le \xi$): First, the UAV hovers at the location $(-\hat{x},H)$ for time $t \in  [0,T/2-\hat{x}/V]$; next, the UAV flies from $(-\hat{x},H)$ to $(\hat{x},H)$ with the maximum speed $V$ during the time interval $t \in (T/2-\hat{x}/V,T/2+\hat{x}/V)$; finally, the UAV hovers at the location $(\hat{x},H)$ for remaining time $t \in [T/2+\hat{x}/V,T]$. In other words, the optimal UAV trajectory is
\begin{align}\label{eqn:hovering:flying:hovering}
x^{\star\star}(t) =
\left\{
\begin{array}{ll}
-\hat{x}, &t \in [0,T/2-\hat{x}/V]\\
Vt -VT/2,& t \in  (T/2-\hat{x}/V,T/2+\hat{x}/V)\\
\hat{x},&t \in [T/2+\hat{x}/V,T].
\end{array}
\right.
\end{align}Here, we have $\hat{x} = \xi$ if $T \ge 2\xi/V$; while $\hat{x} = VT/2$ if $T < 2\xi/V$. Accordingly, $E^{\star\star} = \int_{0}^T Q_1(x^{\star\star}(t))\mathrm{d} t + \int_{0}^T Q_2(x^{\star\star}(t))\mathrm{d} t$.
\end{proposition}
\begin{IEEEproof}
See Appendix \ref{appendix:C}.
\end{IEEEproof}

Proposition \ref{proposition:4} provides key insights on how to maximize the equal energy transferred to the two ERs when the two ERs are distributed far apart with $D > 2H/\sqrt{3}$. If the charging duration is long (i.e., $T > 2\xi/V$), then the UAV should hover at two symmetric locations $(-\xi,H)$ and $(\xi,H)$ with equal time and travel from one location to the other with the maximum speed. Otherwise, if the charging duration is short (i.e., $T \le 2\xi/V$), then the UAV should keep flying at its maximum speed from one ER to the other by following a symmetric trajectory around their middle point $(0, H)$, without hovering over any of them due to the insufficient charging time.

\subsection{Hover-Fly-Hover Trajectory Design for (P1) when $D > 2H/\sqrt{3}$, $\lambda_1^\star = \lambda_2^\star$, and $\alpha_1 \neq \alpha_2$}


Last, we consider the case when $D > 2H/\sqrt{3}$, $\lambda_1^\star = \lambda_2^\star$, and $\alpha_1 \neq \alpha_2$. In this case, we propose an efficient hover-fly-hover design based on the optimal solution to (P3) in Proposition \ref{proposition:4}. In particular, denote the two hovering locations as $\hat{x}_1$ and $\hat{x}_2$, where we have $ - \xi \le \hat{x}_1 \le  \xi$ and $\hat{x}_1 \le \hat{x}_2 \le  \xi$. Then, the proposed hover-fly-hover design is described as follows: First, the UAV hovers at the location $(\hat{x}_1,H)$ for the time interval $t \in \hat{\mathcal T}_1 \triangleq [0,\hat{t}]$; then, the UAV flies from $(\hat{x}_1,H)$ to $(\hat{x}_2,H)$ with the maximum speed $V$ for $t \in \hat{\mathcal T}_2 \triangleq(\hat{t},\hat{t} + (\hat{x}_2-\hat{x}_1)/V)$; finally, the UAV hovers at the location $(\hat{x}_2,H)$ for $t \in \hat{\mathcal T}_3 \triangleq [\hat{t} + (\hat{x}_2-\hat{x}_1)/V,T]$. In other words, we have
\begin{align}\label{eqn:hovering:flying:hovering:2}
\hat x(t) =
\left\{
\begin{array}{ll}
\hat{x}_1, &t \in \hat{\mathcal T}_1\\
\hat{x}_1 + V(t -\hat{t}),& t \in \hat{\mathcal T}_2\\
\hat{x}_2,& t \in \hat{\mathcal T}_3.
\end{array}
\right.
\end{align}
Here, the two hovering locations $\hat{x}_1$ and $\hat{x}_2$ as well as the hovering duration $\hat{t}$ are design variables. Accordingly, the transferred energy to each ER $k\in\{1,2\}$ is given by
\begin{align*}
& \hat E_k(\hat{x}_1,\hat{x}_2,\hat{t}) = \int_{0}^T Q_k(\hat{x}(t))\mathrm{d} t \nonumber\\
= &\hat t Q_k(\hat x_1) + (T-\hat{\tau} - (\hat{x}_2-\hat{x}_1)/V) Q_k(\hat x_2)  \nonumber\\
&+ \frac{\eta\beta_0 P}{VH} \left( \arctan\left(\frac{\hat x_2-x_k}{H}\right) - \arctan\left(\frac{\hat x_1-x_k}{H}\right) \right),
\end{align*}
and the total energy harvested by the two ERs is $\hat E(\hat{x}_1,\hat{x}_2,\hat{t}) = \hat E_1(\hat{x}_1,\hat{x}_2,\hat{t}) + \hat E_2(\hat{x}_1,\hat{x}_2,\hat{t})$.

To find the optimal $\hat{x}_1$, $\hat{x}_2$, and $\hat{t}$, we employ a two-dimensional search over $\hat{x}_1 \in [-\xi,\xi]$ and $\hat{x}_2 \in[\hat{x}_1,\xi]$. Under each given pair of $\hat{x}_1$ and $\hat{x}_2$, $\hat{t}$ should be chosen such that $\hat E_1(\hat{x}_1,\hat{x}_2,\hat{t})/\hat E_2(\hat{x}_1,\hat{x}_2,\hat{t}) = \alpha_1/\alpha_2$. By comparing the values of $\hat E(\hat{x}_1,\hat{x}_2,\hat{t})$, we can find the variables $\hat{x}_1$, $\hat{x}_2$, and $\hat{t}$ achieving the maximum total transferred energy, which are denoted as $\hat{x}^{\star\star}_1,\hat{x}^{\star\star}_2$, and $\hat{t}^{\star\star}$, respectively. By substituting them in (\ref{eqn:hovering:flying:hovering:2}), an efficient solution $\{x^{\star\star}(t)\}$ and the corresponding $E^{\star\star}$ to (P1) are finally obtained.

\begin{remark}
Although it is difficult to rigorously prove its optimality, we expect that such a hover-fly-hover design is quite efficient for problem (P1). For example, when the charging duration $T$ becomes sufficiently large, the objective value of (P1) obtained by the hover-fly-hover design can be shown to converge to the optimal value of (P2) asymptotically as the effect of the finite traveling time between the two ERs diminishes as $T \to \infty$.
\end{remark}

\section{Numerical Results}

In this section, we present numerical results to evaluate the performance of UAV-enabled WPT in a two-user setup. The parameters are set to be the same as for Example \ref{example:1}. First, Figs. \ref{fig:3D:1} and \ref{fig:3D:2} show the (normalized) energy region when $D = 5$ m $< 2H/\sqrt{3}$ and $D = 8$ m $> 2H/\sqrt{3}$, respectively. It is observed from Fig. \ref{fig:3D:1} that when $D = 5$ m, the Pareto boundary of the energy region does not depend on the charging duration $T$ and the UAV's speed constraint $V$. This is because in this case, the optimal solution to problem (P1) is the same as that to (P2), in which the UAV hovers at a fixed location during the whole charging period (see Proposition \ref{proposition:3}). It is observed from Fig. \ref{fig:3D:2} that when $D = 8$ m, the achievable energy region under finite $T$ and $V$ is a non-convex set in general, and it is smaller than that under infinite $T$ and/or $V$ (i.e., with the UAV speed constraints in (\ref{eqn:con:2}) ignored). Furthermore, the case with $T = 1$ s and $V  =10$ m/s achieves a larger energy region than that of the benchmark case with $V= 0$, which is obtained by varying the UAV location above the line between the two ERs. This shows that the UAV mobility is beneficial for enlarging the energy region for ERs.

\begin{figure}
\centering
 \epsfxsize=1\linewidth
    \includegraphics[width=7cm]{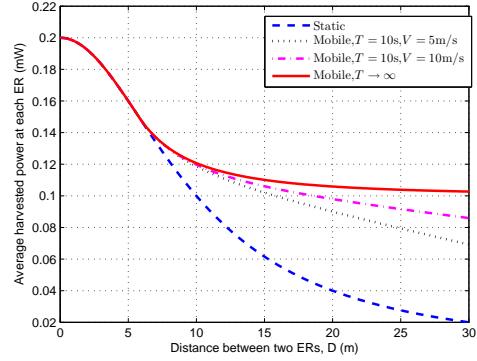}
\caption{Average harvested power at each ER versus the distance $D$ between the two ERs.} \label{fig:3D:3}\vspace{-2em}
\end{figure}

Fig. \ref{fig:3D:3} shows the common average harvested power at each ER in the case with $\alpha_1 = \alpha_2 = 1/2$ versus the distance $D$ between the two ERs. It is observed that when $D \le 2H/\sqrt{3} = 5.77{\text{m}}$, the static UAV design (or equivalently with $V=0$) achieves the same performance as the mobile UAV design, regardless of the speed constraint $V$ of the UAV and the charging duration $T$. By contrast, when $D > 5.77$m, the proposed mobile UAV design achieves larger average harvested energy than the static UAV design (with the UAV fixed at $(0,H)$), and the performance gain becomes more pronounced as $D$ increases. Furthermore, it is observed that as the UAV's speed limit $V$ becomes large, the average harvested power by the ERs is closer to the upper bound with $T\to\infty$ or $V\to\infty$, as the traveling time between the two ERs becomes insignificant.

\vspace{-1em}

\section{Conclusion}

This paper studied a two-user UAV-enabled WPT system. We characterized the Pareto boundary of the achievable energy region of the two ERs by optimizing the UAV's trajectory under the maximum speed constraints. It was shown that when the distance $D$ between the two ERs is no larger than $2/\sqrt{3}$ times of the UAV's flying altitude $H$ (i.e., $D \le 2H/\sqrt{3}$), each Pareto-boundary energy pair is achieved when the UAV hovers above a fixed location between them throughout the charging period, and the energy region is a convex set. On the other hand, when $D > 2H/\sqrt{3}$, the Pareto-boundary of the energy region is generally achieved when the UAV follows a hover-fly-hover trajectory, where the UAV hovers at two different locations in a time-sharing manner, and flies from one location to the other with the maximum speed. It was shown by simulations that the proposed UAV-enabled WPT with the controllable mobile ET is a cost-effective solution to improve the performance of the conventional WPT with fixed ET. In future work, we will extend our study to the general scenario with more than two ERs.

\appendix

\subsection{Proof of Lemma \ref{lemma:3.2}}\label{appendix:A}

In the case with $\lambda_1 = \lambda_2 = \lambda$, the first-order derivative $\psi_{\lambda_1,\lambda_2}'(x(t))$ in (\ref{eqn:first:order}) can be simplified as\vspace{-1em}

\begin{small}\begin{align*}
  &\psi_{\lambda,\lambda}'(x(t)) = -4\eta\beta_0 P \lambda x(t) \times\\
    &\frac{x^4(t) + 2(D^2/4+H^2)x^2(t) - 3/16D^4 + H^4-H^2D^2/2}{(x^2(t) + D^2/4 + H^2 -Dx(t))^2(x^2(t) + D^2/4 + H^2 + Dx(t))^2}.
\end{align*}
\end{small}Then we consider the two cases with $D > 2H/\sqrt{3}$ and $D \le 2H/\sqrt{3}$, respectively.

In the first case with $D > 2H/\sqrt{3}$, it follows that
\begin{align}
  \psi_{\lambda,\lambda}'(x(t))\left\{
  \begin{array}{ll}
  > 0, & {\text{if}}~x(t)\in [-D/2,-\xi)\\
  \le 0, & {\text{if}}~x(t)\in [-\xi,0] \\
  >  0, & {\text{if}}~x(t)\in (0,\xi) \\
  \le 0, & {\text{if}}~x(t)\in [\xi,D/2].
  \end{array}
  \right.
\end{align}
Note that we have $\xi < D/2$ here, provided that $H > 0$. We can accordingly conclude that $\psi_{\lambda,\lambda}(x(t))$ is monotonically increasing, decreasing, increasing, and decreasing over $x(t)\in [-D/2,-\xi)$, $[-\xi,0]$, $(0,\xi)$, and $[\xi,D/2]$, respectively. Note that $\psi_{\lambda,\lambda}(-\xi) = \psi_{\lambda,\lambda}(\xi)$. Then both $x^*_{\lambda,\lambda}(t) = -\xi$ and $x^*_{\lambda,\lambda}(t) = \xi$ are the optimal solution to problem (\ref{eqn:sub:1}).

In the second case with $D \le 2H/\sqrt{3}$, we have
\begin{align}
  \psi_{\lambda,\lambda}'(x(t))\left\{
  \begin{array}{ll}
  > 0, & {\text{if}}~x(t)\in [-D/2,0)\\
  \le 0, & {\text{if}}~x(t)\in [0,D/2].
  \end{array}
  \right.
\end{align}
In this case, $x^*_{\lambda,\lambda}(t) = 0$ is the optimal solution to problem (\ref{eqn:sub:1}).

\subsection{Proof of Proposition \ref{proposition:1}}\label{appendix:B}
In the case with $D \le 2H/\sqrt{3}$, this proposition directly follows from Lemma \ref{lemma:3.2}, by noting that the optimal solution $x^*_{\lambda^\star,\lambda^\star}(t) = 0$ to problem (\ref{eqn:sub:1}) is indeed unique, $\forall t\in\mathcal T$. Substituting this into \eqref{eqn:harvested:energy}, we have the energy transferred to each ER $k\in\{1,2\}$ to be identical as $E_k^\star = E_k(\{x^*_{\lambda^\star,\lambda^\star}\}) = TQ_k(0)$. Furthermore, note that the constraints in \eqref{eqn:con:1} are met with equality at the optimality, and hence we have $\frac{\alpha_1}{\alpha_2} = 1$ and accordingly $\alpha_1 = \alpha_2$ in this case.

In the other case with $D > 2H/\sqrt{3}$, problem (\ref{eqn:sub:1}) under $\lambda_1^\star = \lambda_2^\star = \lambda^\star$ has two optimal solutions $x^*_{\lambda^\star,\lambda^\star} = -\xi$ and $x^*_{\lambda^\star,\lambda^\star} = \xi$. In this case, the optimal solution to (P2) should time-share between the two solutions such that we have $x^\star(t) = -\xi$ and $x^\star(t) = \xi$ for durations $\tau$ and $T-\tau$, respectively. Furthermore, note that the constraints in \eqref{eqn:con:1} should be met with equality at the optimality. By combining the above two results, this proposition follows.

\subsection{Proof of Proposition \ref{proposition:4}}\label{appendix:C}

First, it is evident that at the optimal solution to (P3), the harvested energy at the two ERs must be identical. Thus, it is optimal for the UAV to adopt a symmetric trajectory, i.e., if the UAV stays in a location $(-x,H)$ for a certain amount of time, then it will stay in the symmetric location $(x,H)$ for the same duration, since otherwise, one ER will harvest more energy than the other.

Suppose that the UAV stays at two symmetric locations $(-x,H)$ and $(x,H)$ with the same duration with $x \ge 0$. By noting that $Q_1(x) = Q_2(-x)$ and $Q_2(x) = Q_1(-x)$, it is evident that the harvested powers at the two ERs are identical and each can be expressed as $\psi_{\lambda,\lambda}(x)$, with $\psi_{\lambda_1,\lambda_2}(x)$ shown in (\ref{eqn:sub:1}). As shown in Appendix \ref{appendix:A}, $\psi_{\lambda,\lambda}(x)$ is monotonically increasing and decreasing over $x\in [0,\xi)$ and $[\xi,D/2]$, respectively. Therefore, to maximize the transferred energy to the two ERs, the UAV should set $x$ as large as possible provided that $x \le \xi$. In this case, the solution in \eqref{eqn:hovering:flying:hovering} is optimal to problem (P3) subject to the UAV's speed constraints in (\ref{eqn:con:2}). Proposition \ref{proposition:4} thus follows.

\end{document}